\documentclass[epj]{svjour}
\usepackage{graphics}
\usepackage{epsf}
\usepackage{epsfig}
\usepackage{amsmath}
\usepackage{epstopdf}
\usepackage{mathpazo}   
\def\beq{\begin{equation}}
\def\eeq{\end{equation}}
\def\bea{\begin{eqnarray}}
\def\eea{\end{eqnarray}}
\def\beqa{\begin{equation}\begin{array}{l}}
\def\eeqa{\end{array}\end{equation}}
\begin{document}
\title{Determination of two-photon exchange amplitudes from elastic electron-proton scattering data}
\author{Julia Guttmann \inst{1} \and  Nikolai Kivel \inst{1,2,3} 
\and Mehdi Meziane \inst{4} \and Marc Vanderhaeghen \inst{1, 2} 
}
\institute{
Institut f\"ur Kernphysik, Johannes Gutenberg-Universit\"at, D-55099 Mainz, Germany
\and 
Helmholtz Institut Mainz, Johannes Gutenberg-Universit\"at, D-55099 Mainz, Germany
\and 
Petersburg Nuclear Physics Institute, Gatchina, St. Petersburg 188350, Russia
\and 
Physics Department, College of William and Mary,
Williamsburg, VA 23187, USA
}
\date{Received: date / Revised version: date}
\abstract{
Using the available cross section and polarization data for  elastic electron-proton scattering, 
we provide an extraction of the two-photon exchange amplitudes at a common value of four-momentum transfer, around $Q^2 = 2.5$~GeV$^2$. 
This analysis also predicts the 
$e^+ p / e^- p$ elastic scattering cross section ratio, which will be measured by forthcoming experiments. 
 % end of PACS codes
} %end of abstract
\authorrunning{J. Guttmann \it {et al.}}
\maketitle
The electromagnetic form factors (FFs) of the nucleon have been explored extensively during the last 
50 years with ever increasing accuracy. 
The tool to extract the electromagnetic FFs is provided by the one-photon ($1 \gamma$) exchange approximation to elastic electron-nucleon scattering. Precision measurements of the proton electric to magnetic FF ratio at larger momentum transfers using polarization experiments~\cite{Jones00,Punjabi:2005wq,Gayou02,Puckett:2010ac} have revealed significant discrepancies in recent years  with unpolarized experiments using the Rosenbluth 
technique~\cite{Andivahis:1994rq,Christy:2004rc,Qattan:2004ht}, when analyzing both within the $1 \gamma$-exchange framework. 
As both techniques were scrutinized and their findings confirmed by different experimental groups and using different set-ups, it became clear that some other explanation was needed to explain the difference. 
It was proposed that two-photon ($2 \gamma$) exchange processes are the most likely 
explanation of this difference~\cite{Guichon:2003qm,Blunden:2003sp,Chen:2004tw}. 
Their study has  received a lot of attention lately, see~\cite{Carlson:2007sp} for a recent review (and references therein), and~\cite{Arrington:2007ux} for a global analysis of elastic $ep$ scattering including a model for the $2 \gamma$ corrections. 

In recent years dedicated experiments were performed to test the Rosenbluth method~\cite{Qattan:2004ht} and to measure the $2 \gamma$-corrections to the polarization observables~\cite{Meziane:2010}. Using the theoretical framework to describe elastic $ep$ scattering beyond 
the $1 \gamma$-approximation, as laid out in Ref.~\cite{Guichon:2003qm}, 
both experiments allow for the first time to provide an empirical determination of the 
$2 \gamma$-amplitudes at a common value of the momentum transfer, $Q^2 = 2.64$~GeV$^2$. 
We extract the resulting $2 \gamma$-amplitudes in this work and provide predictions for 
forthcoming experiments. 

To describe the process, 
$l(k)+N(p)\rightarrow l(k')+N(p')$, 
we adopt the definitions~:
$P=(p+p')/2$, $K=(k+k')/2$, $q=k-k'$,
and choose 
$Q^{2}=-q^{2}$ and $\nu =K \cdot P$ 
as the independent kinematical invariants.  
Neglecting the electron mass, the elastic $ep$ scattering amplitude 
can be expressed through 3 independent structures~\cite{Guichon:2003qm}~:
\begin{eqnarray}
\label{eq:tmatrix}
T_{h, \, \lambda'_N \lambda_N} \,&=&\, 
(e^2 / Q^{2}) \, \bar{u}(k', h)\gamma _{\mu }u(k, h)\,  \nonumber \\
&&\hspace{-1.75cm} \times \, 
\bar{u}(p', \lambda'_N)\left( \tilde{G}_{M}\, \gamma ^{\mu }
-\tilde{F}_{2}\frac{P^{\mu }}{M}
+\tilde{F}_{3}\frac{\gamma \cdot K 
P^{\mu }}{M^{2}}\right) u(p, \lambda_N), \nonumber \\
\end{eqnarray}
with $e$ the proton charge, $M$ its mass, 
$h = \pm 1/2$ the electron helicity and $\lambda_N$ 
($\lambda'_N$) the incoming (outgoing) proton helicities. 
In Eq.~(\ref{eq:tmatrix}), 
\( \tilde{G}_{M},\, \tilde{F}_{2},\, \tilde{F}_{3} \) are 
complex functions of \( \nu  \) and \( Q^{2} \). 
For convenience, we also use
$\tilde{G}_{E}\equiv\tilde{G}_{M}-(1+\tau )\tilde{F}_{2}$, 
with $\tau \equiv Q^2 / (2 M)^2$. 
To separately identify the $1 \gamma$- and $2 \gamma$-exchange 
contributions, we
use the decomposition $\tilde G_M = G_M + \delta \tilde G_M$, and
$\tilde G_E = G_E + \delta \tilde G_E$, where $G_M$ and $G_E$ are the
usual proton magnetic and electric FFs, which are functions of
$Q^2$ only and are defined from matrix elements of the electromagnetic
current. The amplitudes $ \tilde{F}_{3}$, $\delta
\tilde{G}_{M}$, and $\delta \tilde G_E$, originate from processes
involving the exchange of at least two photons, and are of order $e^2$.
When calculating observables beyond the $1 \gamma$-approximation, 
it is convenient to  
express the real parts ( ${\cal R} $ ) of the $2 \gamma$-amplitudes relative to the magnetic FF as~:
\begin{eqnarray}
&&Y_M \equiv {\cal R} \left( \delta \tilde G_M / G_M \right), 
\quad
Y_E \equiv {\cal R} \left( \delta \tilde G_E / G_M \right), 
\nonumber \\
&&Y_3 \equiv \left( \nu / M^2\right)  {\cal R} \left( \tilde F_3 / G_M \right). 
\end{eqnarray}

The reduced $e^- p \to e^- p$ cross section including the $2 \gamma$-corrections 
becomes~\cite{Guichon:2003qm}
\begin{eqnarray}
\frac{\sigma_R}{G_M^2} 
= 1 + \frac{\varepsilon}{\tau}  \frac{G_E^2}{G_M^2}   
&+& 2 \, Y_M
+ 2 \varepsilon  \frac{G_E}{\tau G_M} \, Y_E \nonumber \\
&+& 2 \varepsilon \left( 1 + \frac{G_E}{\tau G_M} \right) Y_3  
+  {\mathcal{O}}(e^4) ,\;
\label{eq:crossen} 
\end{eqnarray}
where $\varepsilon$ is the virtual photon polarization parameter. 
The corresponding expression for the elastic $e^+ p \to e^+ p$ cross section is obtained 
by changing the sign in front of the $2 \gamma$-amplitudes in Eq.~(\ref{eq:crossen}). 

In the $1 \gamma$-exchange (Born) approximation, the double polarization observables are given by~:
\begin{eqnarray}
P_l^{Born} &=&  	  \sqrt{1 - \varepsilon^2}  \, (2h) \, 
\left( 1 + \frac{\varepsilon}{\tau} \frac{G_E^2}{G_M^2} \right)^{-1}, 
\label{eq:pl_1gamma} \\
P_t^{Born} &=& 									
	- \sqrt{\frac{2\varepsilon (1 -\varepsilon)}{\tau}}  (2h) 
\left( 1 + \frac{\varepsilon}{\tau} \frac{G_E^2}{G_M^2} \right)^{-1} 
	\frac{G_E}{G_M}. 
	\label{eq:pt_1gamma} 
\end{eqnarray}

Including $2 \gamma$-corrections, the polarization transfer ratio can be written 
as~\cite{Guichon:2003qm}:
\begin{eqnarray}
- \sqrt{ \frac{ \tau (1 +\varepsilon)}{2\varepsilon}}  \frac{P_t}{P_l} &=& 						
	\frac{G_E}{ G_M }
      +	 Y_E  - \frac{G_E}{G_M} \, Y_M \nonumber \\
	&+&
	  \left( 1 - \frac{2 \varepsilon}{1 + \varepsilon} \frac{G_E}{G_M} \right) Y_3
	+  {\mathcal{O}}(e^4)  .	\;	\;											
\label{eq:pt_2gamma} 
\end{eqnarray}

For $P_l$ separately, its expression relative to the $1 \gamma$-result of Eq.~(\ref{eq:pl_1gamma}) 
is given by~:
\begin{eqnarray}
\frac{P_l}{P_l^{Born}}  =   1 &-& 2 \varepsilon 
\left( 1 + \frac{\varepsilon}{\tau} \frac{G_E^2}{G_M^2} \right)^{-1} 
\nonumber \\
&\times&
\left\{ 
\left[ \frac{\varepsilon}{1 + \varepsilon} \left( 1 - \frac{G_E^2}{\tau G_M^2} \right) 
+ \frac{G_E}{\tau G_M}  \right] Y_3
\right. \nonumber \\
&&\left. + \frac{G_E}{\tau G_M} \left[  Y_E
 - \frac{G_E}{G_M} Y_M \right] 
 \right\} +  {\mathcal{O}}(e^4).  \; \;
	\label{eq:pl_2gamma}  
\end{eqnarray}
We can make an estimate of the $2 \gamma$-amplitudes from the recent data  
for the $\varepsilon$-dependence of  
  $P_t/P_l$ and $P_l/P_l^{Born}$ at $Q^2 = 2.5$~GeV$^2$, 
  as measured by the JLab/Hall C experiment~\cite{Meziane:2010},  
  and combine them with the precision Rosenbluth measurements of $\sigma_R$ performed at 
  JLab/Hall A~\cite{Qattan:2004ht},  where data exist at a similar value $Q^2 = 2.64$~GeV$^2$. 
  The combination of both experiments 
  allows for the first time to have three observables at a same $Q^2$ value to extract the 
  $\varepsilon$-dependence of the three $2 \gamma$-amplitudes $Y_M$, $Y_E$, and $Y_3$, which are functions of both $Q^2$ and $\varepsilon$.

\begin{figure}
\includegraphics[width=8.5cm]{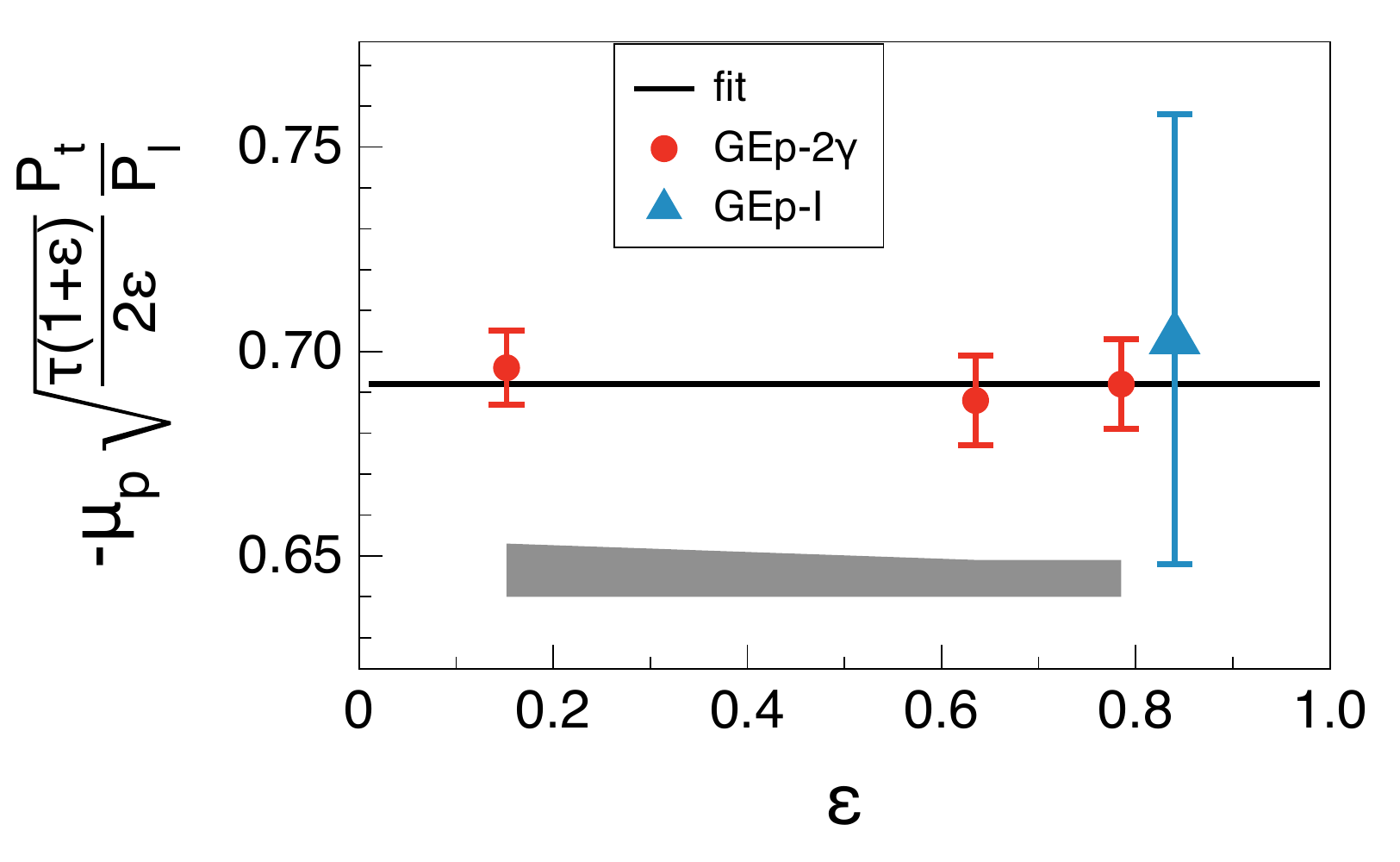}
\caption{ The ratio $- \mu_p \sqrt{\tau (1 + \varepsilon)/(2 \varepsilon)} P_{t}/P_{l}$ as a function of $\varepsilon$ for $Q^{2}=2.5$~GeV$^{2}$. 
The data are from GEp-I (blue triangle)~\cite{Jones00,Punjabi:2005wq}, and  
GEp-$2 \gamma$ (red circles)~\cite{Meziane:2010}~: the error bars (grey band) show the  statistical (systematic) errors respectively. 
The solid curve is an $\varepsilon$-independent fit, given by Eq.~(\ref{eq:ptfit}). 
}
\label{Fig1}
\end{figure}

We start with the data for $P_t / P_l$ as shown on Fig.~\ref{Fig1}. 
It is clearly seen how the data of the dedicated JLab/Hall C GEp-$2 \gamma$ 
experiment~\cite{Meziane:2010} improve on the precision of the 
pioneering GEp-I result~\cite{Jones00,Punjabi:2005wq}. Within their error bars of order 1 \%, the JLab/Hall C experiment does not see any systematic $2 \gamma$-effect on this observable. 
We effectively tried a fit of $- \mu_p \sqrt{  \tau (1 +\varepsilon)/(2\varepsilon)} P_t/P_l$ 
assuming an $\varepsilon$-independent part $R$,
which in the $1 \gamma$-limit equals $\mu_p G_E/G_M$, 
with $\mu_p = 2.793$ the proton magnetic moment, 
supplemented with an $\varepsilon$-dependent part~:
\begin{eqnarray}  
- \mu_p \sqrt{ \frac{\tau (1 +\varepsilon) }{ 2\varepsilon}}  \, P_t / P_l   
= R + B \varepsilon^c \, (1- \varepsilon)^d.  
\label{eq:ptfit0}
\end{eqnarray}
Using a range of values for $c$ and $d$, we found that the value $B$ of 
 is zero within the present error, and the extracted values of $R$ are 
all equal within their error bars. Therefore we conclude that the precision of the present data~\cite{Meziane:2010}  at $Q^2 = 2.5$~GeV$^2$ does not allow to extract an $\varepsilon$-dependent part, in addition to the constant value $R$. 
To extract the $G_E/G_M$ ratio from the constant $R$, we are guided by the Regge limit assumption 
that the $2 \gamma$-correction to $P_t/P_l$ vanishes for $\varepsilon \to 1$. 
We therefore fit $P_{t} / P_{l}$ by its $1 \gamma$-value and obtain at $Q^2 = 2.5$ GeV$^2$~:
\begin{eqnarray}  
R = \mu_p \, G_E / G_M = 0.693 \pm 0.006_{\rm {stat.}} \pm 0.010_{\rm {syst.}}. 
\label{eq:ptfit}
\end{eqnarray}

\begin{figure}
\includegraphics[width=8.5cm]{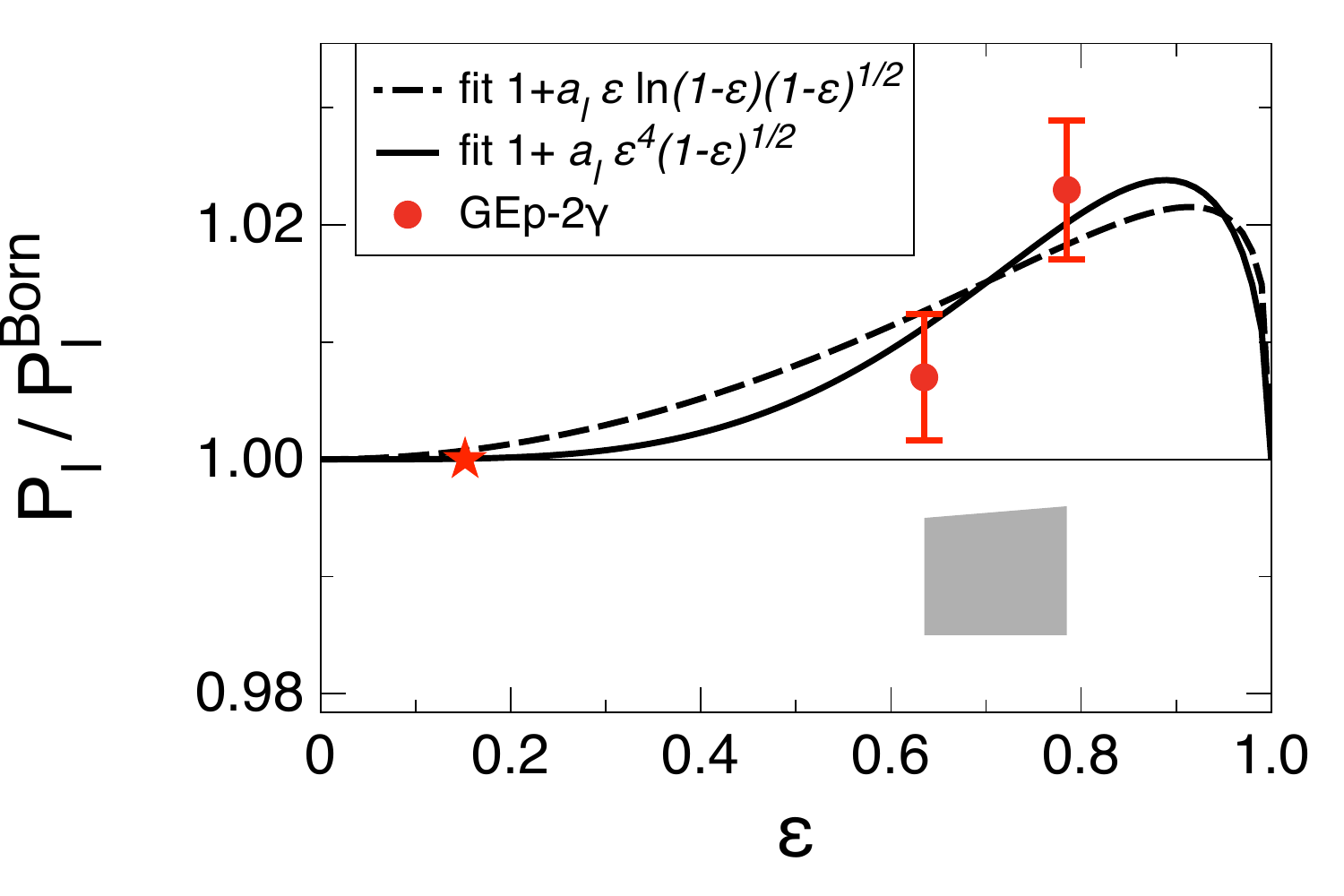}
\caption{ The ratio $P_{l}/P_{l}^{Born}$ as a function of $\varepsilon$ for
$Q^{2}=2.5$~GeV$^{2}$. 
The data points are from the GEp-$2 \gamma$ experiment (red circles)~\cite{Meziane:2010} ~: 
the error bars show the  statistical errors, the grey band shows the systematic errors. 
The star indicates the $\varepsilon$-value at which the data have been normalized to the value 1. 
The curves are two fits described in Eq.~(\ref{eq:plfit})~: Fit 1 (solid curve), Fit 2 (dashed curve). 
}
\label{Fig2}
\end{figure}

To determine the $\varepsilon$-dependence of $P_l$, we conventionally divide by $P_l^{Born}$,  
which is calculated according to Eq.~(\ref{eq:pl_1gamma}), using 
the value of Eq.~(\ref{eq:ptfit}) for $G_E/G_M$.  
Furthermore, the $2 \gamma$-contribution to $P_l$ is expected to be zero in both limits 
$\varepsilon \to 0$ and $\varepsilon \to 1$.  
A perturbative QCD calculation (pQCD)~\cite{Borisyuk:2008db, Kivel:2009eg} gives as limits~: 
 $P_l/P_l^{Born} - 1 \to (1 - \varepsilon)^{1/2}$ for $\varepsilon \to 1$, and 
 $P_l/P_l^{Born} - 1 \to \varepsilon^2$ for $\varepsilon \to 0$. 
 Although the data for  $P_l/P_l^{Born}$ do show a decrease for  $\varepsilon \to 0$,  
 in qualitative agreement with pQCD, the empirical fall-off at $Q^2 = 2.5$~GeV$^2$ is faster than the pQCD prediction.  
 Therefore, we fit  the data for $P_l/P_l^{Born}$ using two different functional forms~:
\begin{eqnarray}
{\rm Fit~1} : \,  P_l / P_l^{Born} &=& 1 + a_l \, \varepsilon^4 (1 - \varepsilon)^{1/2},
\nonumber \\
{\rm Fit~2} : \,  P_l / P_l^{Born} &=& 1 + a_l \, \varepsilon 
\ln(1 - \varepsilon) \, (1 - \varepsilon)^{1/2}.
\label{eq:plfit}
\end{eqnarray}
The fit to the data yields~: 
$a_l = 0.11 \pm 0.03_{\rm {stat.}} \pm 0.06_{\rm {syst.}}$ (Fit 1), 
and  $a_l = -0.032 \pm 0.008_{\rm {stat.}} \pm 0.020_{\rm {syst.}}$ (Fit 2). 

\begin{figure}[h]
\includegraphics[width=8.5cm]{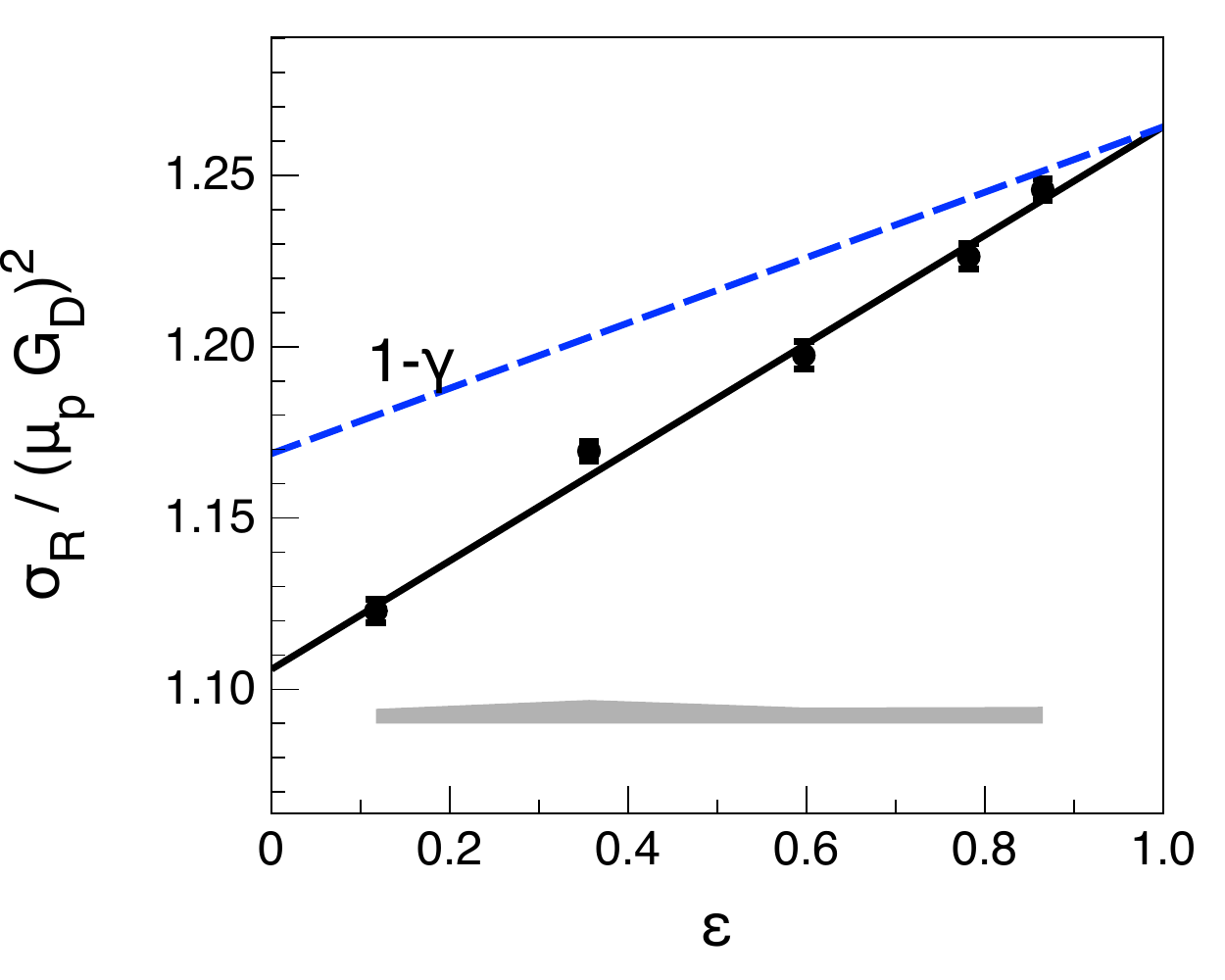}
\caption{ Rosenbluth plots for elastic $e^- p $ scattering:~$\sigma_R$ 
divided by $\mu_p^2  / (1 + Q^2 / 0.71)^{4}$ at $Q^2 = 2.64$~GeV$^2$.  
Solid curve : linear fit to the JLab/Hall A cross section data (circles)~\cite{Qattan:2004ht}. 
Dashed curve : $1 \gamma$-result, using the slope from the polarization data for 
$G_{E}/G_{M}$~\cite{Jones00,Punjabi:2005wq,Gayou02}.   
The grey band shows the systematic errors. 
}
\label{Fig3}
\end{figure}

We next fit the  JLab/Hall A Rosenbluth measurements of $\sigma_R$~\cite{Qattan:2004ht}, 
  shown on Fig.~\ref{Fig3} for $Q^2 = 2.64$~GeV$^2$.  
  The $\sigma_R$ data show a linear behavior in $\varepsilon$, suggesting the fit~:
 \begin{eqnarray}
\sigma_R / (\mu_p G_D)^2 = a + b \, \varepsilon ,  
\label{eq:fitsigma}
\end{eqnarray}
where we follow the convention to factor out the dipole FF $G_D \equiv 1 / (1 + Q^2 / 0.71 \mathrm{GeV}^2)^2$. The fit to the data in Fig.~\ref{Fig3} yields~:
$a = 1.106 \pm 0.006 $ and $b = 0.160 \pm 0.009 $. 

To extract the three $2 \gamma$-amplitudes as well as $G_E/G_M$ and $G_M$ 
from the above three observables at a same $Q^2$ value we have to make two assumptions. 
The first assumption was made in Eq.~(\ref{eq:ptfit0}), where the $\varepsilon$-independent term $R$ 
fixes $G_E/G_M$, see Eq.~(\ref{eq:ptfit}). 
To fix the value of $G_M^2$, we make the assumption, that for 
$\varepsilon \to 1$ (Regge limit) the $2 \gamma$-correction to $\sigma_R$ vanishes, i.e.
$\sigma_R(\varepsilon = 1, Q^2)  =  G_M^2 + G_E^2 / \tau$.
Using the $G_E/G_M$ value extracted from the fit to $P_t/P_l$, and the fitted values of $a$ and $b$ 
entering Eq.~(\ref{eq:fitsigma}), 
then allows to extract the value of $G_M^2$ as~:
\begin{eqnarray}
 \left( \frac{G_M }{ \mu_p G_D} \right)^2 = \frac{a + b }{1 + (1 / \tau) (G_E / G_M)^2}.
 \end{eqnarray}
 For $Q^2 = 2.64$~GeV$^2$, our fit yields~: 
 \begin{eqnarray}
 \left( \frac{G_M }{\mu_p G_D} \right)^2 = 1.168 \pm 0.010 .
 \end{eqnarray}
 
\begin{figure}
\includegraphics[width=8.5cm]{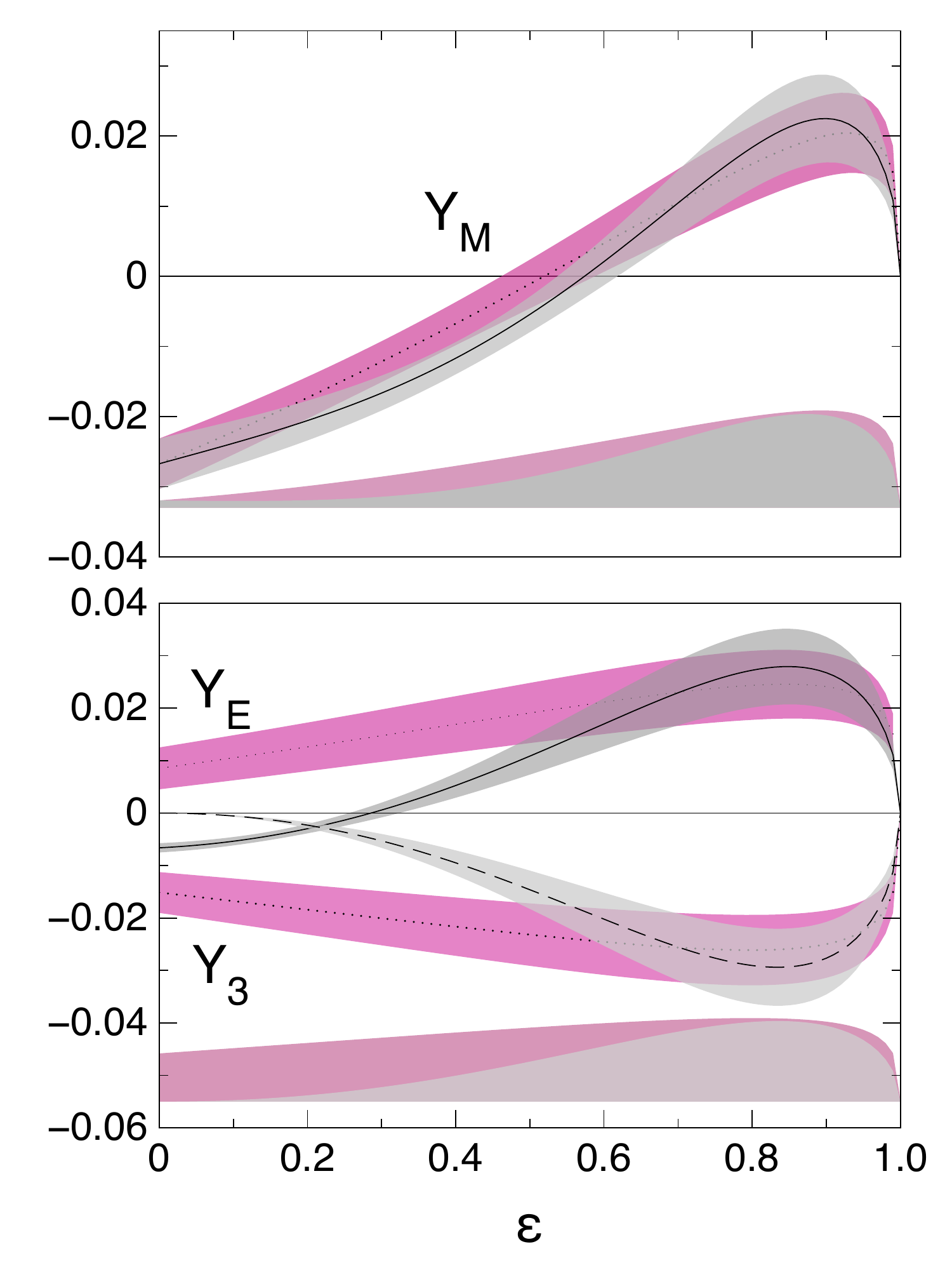}
\vspace{-0.1cm}
\caption{ The extracted $2 \gamma$-amplitudes as a function of $\varepsilon$ for
$Q^{2}=2.64$~GeV$^2$ for the two fits of $P_l$ in Eq.~(\ref{eq:plfit}), with their $1 \sigma$ 
statistical error bands. Fit 1~: grey bands; Fit 2~: red bands. The horizontal bands at the bottom 
of the plots indicate the systematic errors. 
}
\label{Fig4}
\end{figure}

Having specified in  Eqs.~(\ref{eq:crossen}, \ref{eq:pt_2gamma}, \ref{eq:pl_2gamma}) 
the fits of the observables $P_t/P_l$, $P_l / P_l^{Born}$, and $\sigma_R$, 
we next proceed to extract the  $2 \gamma$-amplitudes $Y_M$, $Y_E$, and $Y_3$. The 
 data allow us to perform this analysis at a common value of $Q^2 = 2.64$~GeV$^2$, where we 
 neglect the small difference in $P_t/P_l$ and $P_l/P_l^{Born}$ between their values at $Q^2 = 2.64$~GeV$^2$ and their measured values at $Q^2 = 2.5$~GeV$^2$. 
 In Fig.~\ref{Fig4}, we show the result of this analysis for the $2 \gamma$ amplitudes, 
 together with their $1 \sigma$ error bands. 
 One sees from Fig.~\ref{Fig4} that the amplitude which is best constrained by the available data 
 is $Y_M$. This is because the amplitude $Y_M$ is mainly driven by the $2 \gamma$-effect in the cross section data,  Eq.~(\ref{eq:crossen}), which to good approximation 
 is given by~: $\sigma_R^{2 \gamma} \sim Y_M + \varepsilon \, Y_3$.  
 One notices that the error bands on $Y_M$ resulting from the two different fits 
  of Eq.~(\ref{eq:plfit}) for $P_l$ largely overlap. Except in the large $\varepsilon$ region, 
  the dominance of $Y_M$ by the Rosenbluth data results 
  in its approximate linear rise in $\varepsilon$. 
For $\varepsilon \to 1$, $Y_M$ has to become non-linear provided 
$Y_M + \varepsilon \, Y_3$ remains linear in the limit $\varepsilon \to 1$, 
which we assumed in our present analysis. 
How far the linearity of the Rosenbluth plot extends when approaching $\varepsilon \to 1$ is of course an open question, which will be addressed by the results of a dedicated 
experiment~\cite{JLab_E05_017}.
In contrast to $Y_M$, the amplitudes  $Y_E$ and $Y_3$ are mainly driven by the polarization data. One sees from Fig.~\ref{Fig4} that the error bands overlap in the range ($\varepsilon > 0.6$) where 
data exist for all three observables. In the smaller $\varepsilon$ range, where there is no constraint from the polarization data, one 
sees clear deviations between the two different functional forms for the $\varepsilon$-dependence. We checked that the same conclusion is reached for other forms. 
 One further notices that in the region constrained by the polarization data, the amplitudes $Y_E$ and $Y_3$ are at  the 2-3~\% level and have opposite signs.
 This can be easily understood from Eq.~(\ref{eq:pt_2gamma}) when neglecting the small terms 
 in the $2 \gamma$-amplitudes which are multiplied by $G_E/G_M$. The leading 
 $2 \gamma$-correction to $P_t/P_l$ is to very good approximation proportional to 
 $Y_E + Y_3$. The absence of $2 \gamma$-corrections in the data for 
 $P_t/P_l$ then implies  $Y_E$ and $Y_3$ being of equal magnitude and 
 of opposite sign. Furthermore, the value of $Y_3$ is nearly entirely driven by the data 
 for $P_l$. One indeed obtains from Eq.~(\ref{eq:pl_2gamma}), when again neglecting the 
 small $G_E / G_M$ terms, $P_l^{2 \gamma} \simeq  - 2 \varepsilon^2/(1 + \varepsilon) 
 \, Y_3$, therefore determining $Y_3$.  To improve on the extraction 
 of $Y_E$ and $Y_3$ will require a further improvement in precision of the  polarization experiments.

\begin{figure}[h]
\includegraphics[width=8.5cm]{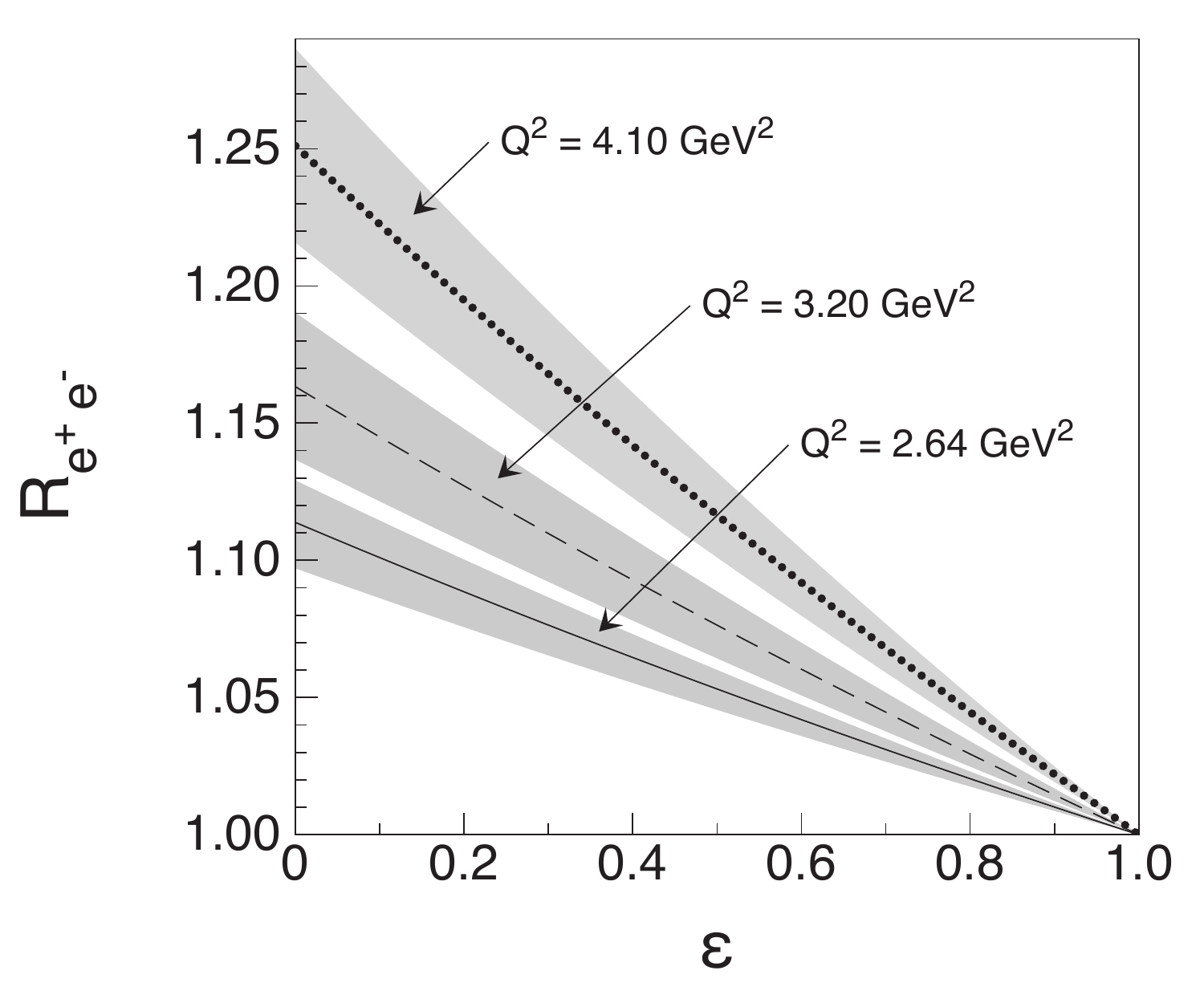}
\caption{ Predictions for the $e^+ p / e^- p$ elastic cross section ratio $R_{e^+ e^-}$ 
as a function of $\varepsilon$, together with their $1 \sigma$ error bands. 
}
\label{Fig5}
\end{figure}

 We next discuss the ratio $R_{e^+ e^-}$ of  $e^+ p / e^- p$ elastic scattering cross sections. The $e^+ p$  elastic scattering observables are obtained from the ones for $e^- p$ by merely changing the  sign in front of the $2 \gamma$-amplitudes. A measurement of the ratio $R_{e^+ e^-}$  therefore provides a test of the $2 \gamma$-amplitudes and is planned in the 
 near future by several experiments~\cite{vepp3,clas,olympus}.  
 In particular, the Olympus Collaboration at DESY plans to measure the $e^+/e^-$ ratio 
 for a value $Q^2 \simeq 2.5$~GeV$^2$~\cite{olympus}.  
Our extraction of the $2 \gamma$-amplitudes at $Q^2 = 2.64$~GeV$^2$ allows to predict the ratio $R_{e^+ e^-}$, which is shown in Fig.~\ref{Fig5}, using Fit 1 in Eq.~(\ref{eq:plfit}). 
 We notice that for $Q^2 = 2.64$~GeV$^2$, $R_{e^+ e^-}$ rises linearly to small $\varepsilon$, 
 reaching $R_{e^+ e^-} = 1.053 \pm 0.004$ for $\varepsilon = 0.5$. 
 In Fig.~\ref{Fig5}, we also show results 
 for the two other values of $Q^2$ where the JLab high-precision Rosenbluth 
 experiment~\cite{Qattan:2004ht} has taken data. At these higher values of $Q^2$, a systematic measurement of the $\varepsilon$-dependence of the polarization observables has not 
 yet been performed. For our analysis of the $Q^2 = 3.2$~GeV$^2$ and  $Q^2 = 4.1$~GeV$^2$ data, 
 we therefore assumed that $P_t / P_l$ can be fitted by its $1 \gamma$-value proportional to 
 $G_E/G_M$ as extracted in~\cite{Jones00,Punjabi:2005wq,Gayou02}. 
 We see from Fig.~\ref{Fig5} that, for a fixed value of $\varepsilon$,  the extracted ratio $R_{e^+ e^-}$ increases with $Q^2$. 
 
 Summarizing, in this work we provided a first combined analysis of high precision Rosenbluth data 
 for elastic electron-proton scattering and recent measurements of the 
 $2 \gamma$-corrections to the polarization observables. As both experiments were performed at 
 a similar $Q^2$ value, we were able to perform an extraction of the three 
 $2 \gamma$-amplitudes  at $Q^2 = 2.64$~GeV$^2$ using empirical results for three observables {\it and} assuming that for $\varepsilon \to 1$ (Regge limit) the $2 \gamma$-amplitudes vanish. We found that one amplitude ($Y_M$) can be reliably extracted from the correction on the unpolarized cross section. The other two amplitudes are at the 2-3 \% level and of opposite sign, partially cancelling each other in the polarization transfer ratio. 
 Our extraction allowed us to provide a prediction 
 of the $e^+ p / e^- p$ elastic cross section ratio, for which dedicated measurements by the Olympus@DESY experiment are underway. Over the measured range of this experiment, the $2 \gamma$-corrections to the  $e^+ p / e^- p$ elastic cross section ratio 
 are predicted to vary in the 1 - 6 \% range.

 \section*{Acknowledgments} 
The work of J.G. was supported by the Research Centre ``Elementarkraefte
und Mathematische Grundlagen" at the University
Mainz. The authors like to thank 
J. Arrington, R. Milner, L. Pentchev, and C. Perdrisat  
for helpful communications and discussions.

\end{document}